\documentclass[aps,preprint,superscriptaddress]{revtex4}

\usepackage{amsmath}
\usepackage{amssymb}
\usepackage{graphicx}
\usepackage{CJK}
\usepackage{keyval}
\usepackage{dcolumn}
\usepackage{bm}
\usepackage{color}
\usepackage{txfonts}
\usepackage[]{sidecap}
\usepackage{comment}
\newcommand{\red}[1]{\textcolor{black}{#1}}

\begin{document}
%\linenumbers
\title{\red{Colossal Band Renormalization and Stoner Ferromagnetism induced by Electron-Antiferromagnetic-Magnon Coupling
}}

\author{T. L. Yu}
\affiliation{Laboratory of Advanced Materials, State Key Laboratory of Surface Physics and Department of Physics, Fudan University, Shanghai 200438, People's Republic of China}

\author{R. Peng}
\affiliation{Laboratory of Advanced Materials, State Key Laboratory of Surface Physics and Department of Physics, Fudan University, Shanghai 200438, People's Republic of China}
\affiliation{Shanghai Research Center for Quantum Sciences, Shanghai 201315, People's Republic of China}

\author{M. Xu}
\author{\red{W. T. Yang}}
\author{\red{Y. H. Song}}
\author{C. H. P. Wen}
\author{Q. Yao}%
\author{X. Lou}
\author{T. Zhang}
\author{W. Li}
\author{X. Y. Wei}
\affiliation{Laboratory of Advanced Materials, State Key Laboratory of Surface Physics and Department of Physics, Fudan University, Shanghai 200438, People's Republic of China}

\author{J. K. Bao}
\author{G. H. Cao}
\affiliation{Department of Physics, Zhejiang University, Hangzhou 310027, People's Republic of China}

\author{P. Dudin}
\affiliation{Diamond Light Source, Harwell Science and Innovation Campus, Didcot OX11 0DE, United Kingdom}

\author{J. D. Denlinger}
\affiliation{Advanced Light Source, 1 Cyclotron Road
Lawrence Berkeley National Laboratory
Berkeley, CA 94720-8229, USA}

\author{V. N. Strocov}
\affiliation{Swiss Light Source, Paul Scherrer Institut, CH-5232 Villigen PSI, Switzerland}

\author{H. C. Xu}\email{xuhaichao@fudan.edu.cn}
\affiliation{Laboratory of Advanced Materials, State Key Laboratory of Surface Physics and Department of Physics, Fudan University, Shanghai 200438, People's Republic of China}

\author{D. L. Feng}\email{dlfeng@fudan.edu.cn}
\affiliation{Laboratory of Advanced Materials, State Key Laboratory of Surface Physics and Department of Physics, Fudan University, Shanghai 200438, People's Republic of China}
\affiliation{Shanghai Research Center for Quantum Sciences, Shanghai 201315, People's Republic of China}
\affiliation{Collaborative Innovation Center of Advanced Microstructures, Nanjing 210093, China}
\affiliation{%
Hefei National Laboratory for Physical Science at Microscale, CAS Center for Excellence in Quantum Information and
Quantum Physics, and Department of Physics, University of Science and Technology of China, Hefei 230026
}%

\begin{abstract}
The interactions between electrons and antiferromagnetic magnons (AFMMs) are  important  for a large class of correlated materials. For example, they  are the most plausible pairing glues in high-temperature superconductors, such as cuprates and iron pnictides.  However,  unlike electron-phonon interactions (EPIs),  clear-cut observations regarding how electron-AFMM interactions (EAIs) affect the band structure are still lacking. Consequently,  critical information on the EAIs, such as its \red{strength} and doping dependence, remains elusive. \red{Here  we directly observe that EAIs induces a kink structure in the band dispersion in Ba$_{1-x}$K$_x$Mn$_2$As$_2$, and subsequently unveil several key characteristics of EAIs.} We found that \red{ the coupling constant} of EAIs can be as large as 6, and it shows huge doping dependence and temperature dependence, all in stark contrast to the behaviors of EPI and beyond our current understanding of EAIs.
Such a colossal \red{renormalization of electronic bands by EAIs}
drives the system to the Stoner criteria, giving the intriguing  ferromagnetic state in Ba$_{1-x}$K$_x$Mn$_2$As$_2$.
Our results expand the current knowledge of EAIs, which may facilitate the further understanding of many correlated materials where EAIs play a critical role, such as high-temperature superconductors.
\end{abstract}

\maketitle

Electron-boson interactions belong to the most fundamental microscopic processes in solids,  which are  responsible for various facinating properties.  For example, electron-phonon interactions (EPIs) could result in conventional superconductivity or charge density waves \cite{PhysRev.108.1175,PhysRevLett.92.086401}, whereas the high temperature superconductivity in cuprate and iron-based superconductors is proposed to be related to the interactions between electrons and
antiferromagnetic (AFM) spin fluctuations, \textsl{i.e.}, magnons  \cite{Science.284:1282, J.Phys.Condens.Matt.16:R755, Sci.Bull.(2016)61(7):561}.
%\red{Anatomy on these interactions, \textit{e.g.}, finding the electron-boson coupling function $g$, is critical for developing corresponding microscopic theories, while the key information lies in their renormalization on the electronic structure.}
The  electron-boson interactions   ``dress" the electrons up, and convert the electrons into quasiparticles. Consequently,
the band structure is \red{renormalized}, and sometimes  an abrupt distortion in the otherwise smooth band  dispersion can be detected by angle resolved photoemission spectroscopy (ARPES) \cite{10.1088/0953-8984/14/24/306,10.1088/1367-2630/11/12/125005}. More specifically, a  kink structure  has been widely observed in  the  dispersion along the nodal direction of many cuprate superconductors \cite{Nature(London)423:398(2003), PRL.87.177007(2001), Nature412:510(2001), PRL.91.157003(2003), PRL.96.117004(2006)}, which is now attributed to the EPI as it is present in the heavily overdoped regime without many AFM fluctuations \cite{PRL.93.117004(2004), PRB79.054528(2009),PhysRevB.69.220502(2004),PhysRevB.73.144507(2006)}.
\red{Theoretically, the effects of electron-boson interaction % is determined by both the coupling function $g$ and the material-specific bosonic structure, and
can be described by a complex self-energy of the electronic structure  \cite{10.1088/0953-8984/14/24/306,10.1088/1367-2630/11/12/125005}.  Based on the ARPES data, one can  extract the self-energy and the Eliashberg function determined by both the electron-boson matrix element $g$ (i.e. interaction potential) and the material-specific bosonic structure \cite{Mahan-book}.
Meanwhile, the electron-boson coupling constant $\lambda$, which characterizes the total renormalization strength and directly gives $T_c$ in BCS theory, is found to be less than 1 for electron-phonon interaction in most metals  \cite{PRL.93.117004(2004),PRB79.054528(2009),BKBO}.}

The magnon effects on the electronic degrees of freedom have been hard to detect in general, either ferromagnetic or antiferromagnetic. In ferromagnets, slight deviations from parabolic dispersion are suspected to be from ferromagnetic magnons but embedded in the electron-electron correlation effect  \cite{10.1103/PhysRevLett.92.097205, Ni110, Fe110}, while a kink well above magnon energy is dominated by the Stoner excitations rather than magnons \cite{https://doi.org/10.1038/s41467-019-08445-1}.
For AFMM, its contribution to the electronic self-energy has not been experimentally identified, although electron-AFMM interactions (EAIs) holds the key to the general understanding of many correlated phenomena like unconventional superconductivity.
%More generally, the electron-magnon interaction has been difficult to separate from the effects of other interactions, such as the electron-electron correlation effect and Stoner excitations in ferromagnets \cite{10.1103/PhysRevLett.92.097205, Ni110, Fe110, https://doi.org/10.1038/s41467-019-08445-1}.
%Slight deviations from parabolic dispersion in ferromagnets are suspected to be from interactions with ferromagnetic magnons \cite{10.1103/PhysRevLett.92.097205, Ni110, 10.1016/j.susc.2007.04.069}, while the corresponding self-energy is embedded in the electron-electron correlation effect. Ferromagnetic magnon may contribute to the band anomaly in Fe, but it is far above the magnon energy and dominated by the Stoner excitation at eV-scale \cite{https://doi.org/10.1038/s41467-019-08445-1}.
The energy scale of AFM spin fluctuations is comparable to the bandwidth in cuprates or iron pnictides, which would lead to a renormalization of the entire band rather than a kink near the Fermi energy ($E_\mathrm{F}$). Therefore, one cannot isolate or identify the EAIs in the retrieved total self-energy.

To reveal the characteristics of the EAIs, one requires a compound that has weak electron correlations and  robust AFMM excitations. This dilemma condition is usually difficult to fulfil, but could be realized in Ba$_{1-x}$K$_x$Mn$_2$As$_2$.
BaMn$_2$As$_2$  is an isostructure material for 122-type Fe-based superconductors \cite{PhysRevLett.101.107006(2003)}. It is an  AFM insulator (AFI)  with a N\'{e}el temperature \emph{T}$_N$ = 625~K. The  magnetic moments of Mn$^{2+}$ ions in BaMn$_2$As$_2$  point along the $c$ axis, forming a G-type AFM order (Fig.~1b). \red{K doping turns Ba$_{1-x}$K$_x$Mn$_2$As$_2$ into a metal. Despite of the carrier doping, the  \emph{T}$_N$ is slightly suppressed and large magnetic moments persist (Fig.~1a) (ref.~\onlinecite{PhysRevLett.108.087005(2012), PhysRevB.85.144523(2012)}).} Inelastic neutron scattering studies find strong AFMM excitations across the phase diagram \cite{PhysRevB.95.224401(2017)}.
The electron-electron correlations are weak according to previous studies on BaMn$_2$As$_2$ (ref.~\onlinecite{PhysRevB.94.155155(2016)}).
\red{At $x>0.19$, it shows a novel ferromagnetic ground state (FM) with the FM moments in the $ab$-plane contributed by As-$4p$ orbital rather than the canting of Mn AFM moments \cite{PhysRevB.85.144523(2012),PhysRevLett.114.217001(2015)},
%which is speculated to be a itinerant ferromagnetism that approaches a half metal state at $x=0.4$ (ref.~\onlinecite{PhysRevLett.111.047001(2013)}),
while the underlying mechanism of the emergent itinerant FM is unknown.
%At low temperatures, ferromagnetism (FM) arises in the $ab$-plane due to As-$4p$ states according to the X-ray dichroism data, while the Mn moments do not contribute to the ferromagnetism \cite{PhysRevB.85.144523(2012),PhysRevLett.114.217001(2015)}.  This ferromagnetic state was speculated to be a half-metal; however, direct evidence is still lacking \cite{PhysRevLett.111.047001(2013)}.
If the weak  electron-electron correlations remains in Ba$_{1-x}$K$_x$Mn$_2$As$_2$ following the parent compound BaMn$_2$As$_2$ \cite{PhysRevB.94.155155(2016)},  it would be an ideal playground for studying EAIs, as its AFMM excitations should be  strong.}

\begin{figure}[b]
    \centering
    \includegraphics[width=160mm]{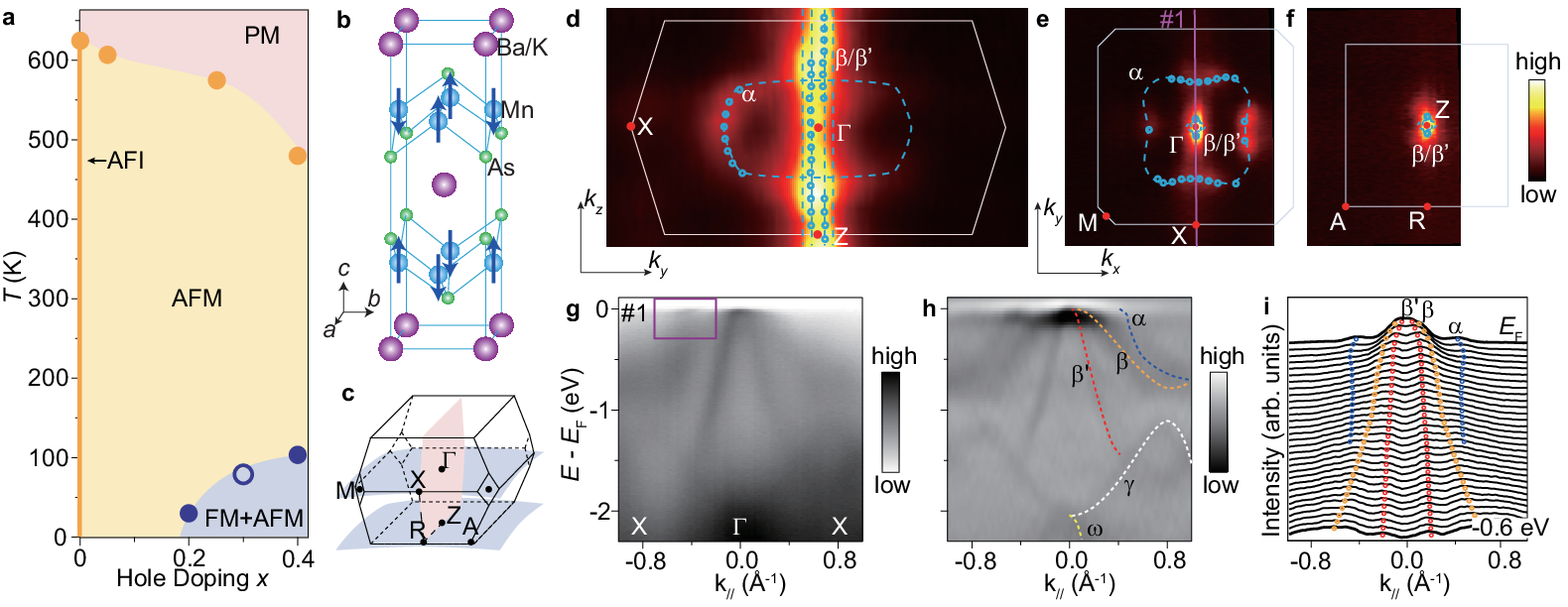}
	\caption{\textbf{Phase diagram and the basic electronic structure of Ba$_{1-x}$K$_x$Mn$_2$As$_2$ ($\mathbf{x}$ = 0.3).} (\textbf{a}) Phase diagram of Ba$_{1-x}$K$_x$Mn$_2$As$_2$. The open circle indicates the Curie temperature determined by magnetic susceptibility measurements on our sample. The filled circles indicate data from ref.~\onlinecite{PhysRevB.85.144523(2012),PhysRevB.80.100403(2009),PhysRevB.87.144418(2013),PhysRevLett.111.047001(2013)}. (\textbf{b-c}) Crystal structure of Ba$_{1-x}$K$_x$Mn$_2$As$_2$ and the corresponding three-dimensional Brillouin zone. The blue arrows indicate the alignment of the Mn magnetic moments in the G-type antiferromagnetic (AFM) state. (\textbf{d-f}) Photoemission intensity maps in the high symmetric planes of the Brillouin zone, integrated over an energy window of \emph{E$_{F}$} $\pm$ 20 meV. The $k_y-k_z$ map in the $\Gamma$XZ plane is measured with photon energies ranging from 58 eV to 100 eV, while the $k_x-k_y$ maps in the $\Gamma$XM plane and ZRA plane are measured with 78 eV and 62 eV photons, respectively. (\textbf{g-h}) Photoemission intensity and its second derivative with respect to energy along cut \#1 in panel \textbf{e}. The bands are outlined by the dashed curves.  (\textbf{i}) Momentum distribution curves (MDCs) along cut \#1. The circles track the local maxima for the dispersion of bands $\alpha$, $\beta$ and $\beta$'. The ARPES data were measured at 30~K.}
\end{figure}

\section*{Weak electron correlations in Ba$_{1-x}$K$_x$Mn$_2$As$_2$}
Figures~1d-f show ARPES results measured using vacuum ultra-violet (VUV) photons, which resolve the Fermi surface structure of Ba$_{1-x}$K$_x$Mn$_2$As$_2$ ($x$ = 0.3) in its three-dimensional Brillouin zone. In the $\Gamma$-X-M plane measured with 78~eV photons, the Fermi surfaces consist of a large pocket $\alpha$  and two small elliptical pockets $\beta$ and $\beta$' perpendicular to each other, all centred at $\Gamma$. The $\alpha$ band is absent in the ZRA plane and highly dispersive along $k_z$, forming a drum-shaped Fermi surface in the $k_y-k_z$ plane. $\beta$ and $\beta$' are two-dimensional  with negligible dispersion along $k_z$. Except for some variation in the spectral weight due to the photoemission matrix element, the Fermi surfaces measured using VUV photons are consistent with
the soft X-ray ARPES results (Supplementary Section~1), confirming the bulk nature of the measured bands. \red{The single set of bands and Fermi surface, and the fact that the Fermi surface volume is consistent with its doping level demonstrate that   Ba$_{1-x}$K$_x$Mn$_2$As$_2$ samples are homogeneous (Supplementary Section~2, \cite{PhysRevB.94.155155(2016), nphys1908, PhysRevB.87.144418(2013)}).}

Along $\Gamma$-X, the bands $\alpha$, $\beta$, and $\beta$' that cross $E_\mathrm{F}$ are all resolved, showing hole-like dispersions (Figs.~1g-i). Both the Fermi surface structure and band dispersion of Ba$_{1-x}$K$_x$Mn$_2$As$_2$ roughly agree with a rigid band shift of the pristine BaMn$_2$As$_2$ \cite{PhysRevLett.108.087005(2012), PhysRevB.94.155155(2016)}.
The overall bandwidths  near $E_\mathrm{F}$ show little renormalization compared with the DFT calculations of BaMn$_2$As$_2$ (ref.~\onlinecite{PhysRevB.94.155155(2016), PhysRevB.85.144523(2012)}), consistent with the weakly correlated As-$4p$ states. The weak band renormalization in Ba$_{1-x}$K$_x$Mn$_2$As$_2$ is distinct from the strongly renormalized bands in the isostructural Ba$_{1-x}$K$_x$Fe$_2$As$_2$ (ref.~\onlinecite{PhysRevLett.105.117003(2010)}). The weak electron-electron correlation would contribute little electronic self-energy near $E_\mathrm{F}$, which provides a clean playground for studying electron-boson interactions in Ba$_{1-x}$K$_x$Mn$_2$As$_2$.

\section*{Strong electron-magnon interactions in Ba$_{1-x}$K$_x$Mn$_2$As$_2$}

\begin{figure}[hb]
	\centering
	\includegraphics[width=140mm]{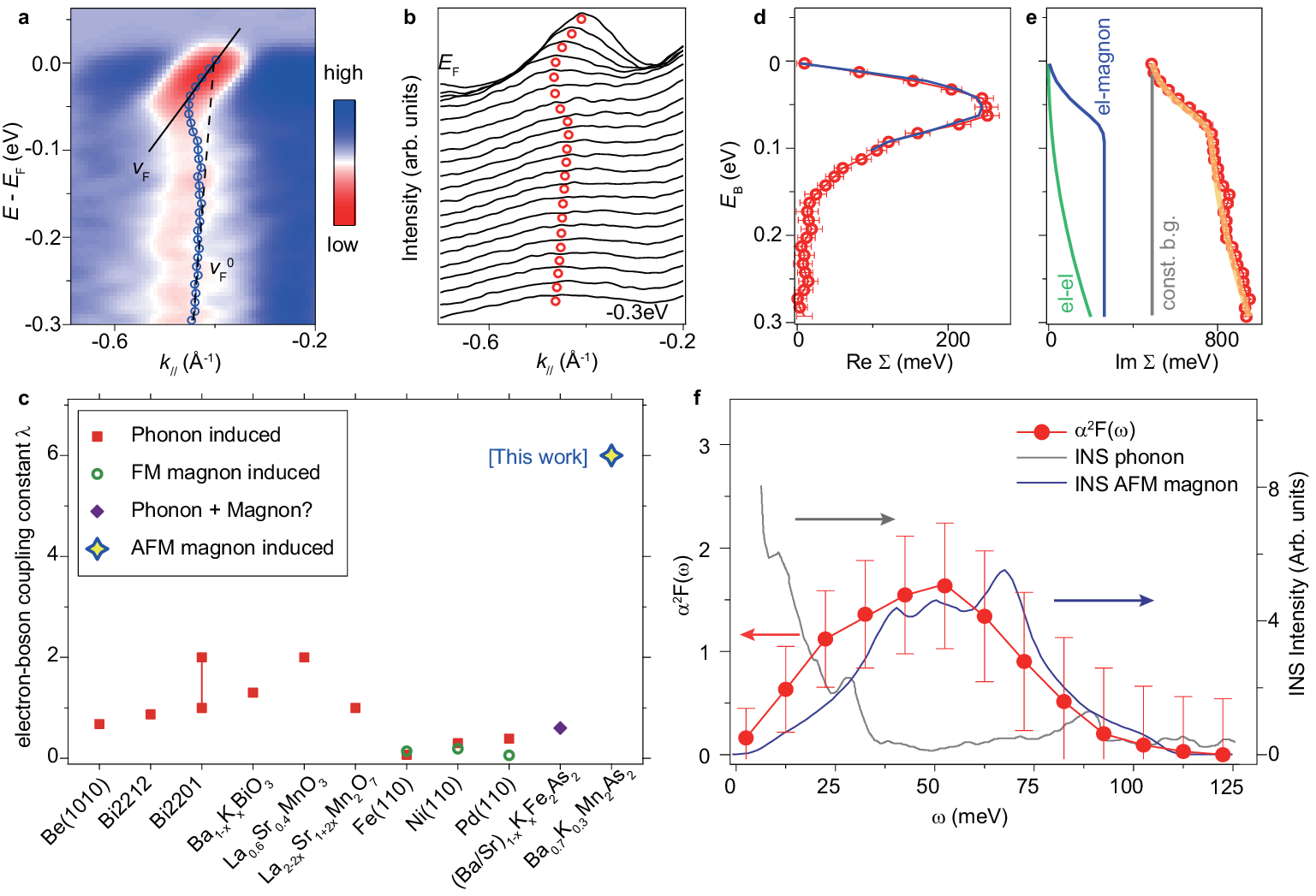}
	\caption{\textbf{Kink and self-energy analysis showing the electron-magnon interaction for Ba$_{0.7}$K$_{0.3}$Mn$_2$As$_2$.} (\textbf{a}) Second derivative spectra of the enlargement near the Fermi crossing of the $\alpha$ band (rectangular area in Fig.~1g), overlaid with the fitting of the Fermi velocity \emph{v$_{F}$} (solid line) and the bare band velocity \emph{v$_{F}^0$} (dashed line). (\textbf{b}) Momentum distribution curves (MDCs) of photoemission intensity around the Fermi crossing of the $\alpha$ band. The circles track the MDC peaks and illustrate the band dispersion. \red{(\textbf{c}) The reported electron-boson coupling constant obtained based on MDC analysis of  kinks in ARPES data of Be(1010) \cite{Be1010}, Bi2212 \cite{Bi2212}, Bi2201 \cite{Bi2201}, Ba$_{1-x}$K$_x$BiO$_3$ \cite{BKBO}, La$_{0.6}$Sr$_{0.4}$MnO$_3$ \cite{LSMO113}, La$_{2-2x}$Sr$_{1+2x}$Mn$_2$O$_7$ \cite{LSMO327}, Fe(110) \cite{Fe110,Fe110/2}, Ni(110) \cite{Ni110}, Pd(110) \cite{Pd110}, and (Ba/Sr)$_{1-x}$K$_{x}$Fe$_2$As$_2$ \cite{BKFA}, together with that of Ba$_{0.7}$K$_{0.3}$Mn$_2$As$_2$.}
(\textbf{d}) Re~$\Sigma$ (red) and the KK transformation of Im~$\Sigma$ below 115~meV (blue curve). (\textbf{e}) The KK transformation of Re~$\Sigma$ (blue curve), the electron-electron scattering (green curve), and a constant background (grey curve). The orange curve shows their combination, which matches well with Im~$\Sigma$ (red circles). \red{(\textbf{f}) Eliashberg function (red) extracted from the Im~$\Sigma$, compared with the phonon (purple) and magnon (blue circles) intensity from inelastic neutron scattering (INS) and simulated magnetic intensity (blue curve) on Ba$_{1-x}$K$_x$Mn$_2$As$_2$ ($x$=0.25) (ref.~\onlinecite{PhysRevB.95.224401(2017)}).}
	}
\end{figure}

Band $\alpha$ shows a kink around the binding energy $E_B$=50~meV, which is better visualized in the second derivative image and MDCs (Figs.~2a-b). This kink in electronic dispersion is a general signature of electron-boson interactions \cite{10.1088/1367-2630/11/12/125005}.
The Fermi velocity $\emph{v}_F$ is reduced from the bare band velocity $\emph{v}^0_F$ by a factor of $\sim$7 (Fig.~2a), from which one can estimate the electron-boson coupling \red{constant} $\lambda$=$\emph{v}_{F}^{0}$/$\emph{v}_{F}$-1 to be as large as 6.
\red{The observed electron-boson coupling \red{constant} is colossal as compared to those relating with electron ferromagnetic magnons interations and EPI (Fig.~2c) \cite{Be1010,Bi2212,Bi2201,BKBO,LSMO113,LSMO327,Fe110,Fe110/2,Ni110,Pd110,BKFA}%,LSMO227over}., which is typically less than unity in transition-metal ferromagnets ($\lambda$ = 0.19 in Ni for instance\cite{Ni110}).
}
%It is also much larger than the EPI coupling \red{constant} found in most metals.
%Even for Ba$_{0.51}$K$_{0.49}$BiO$_3$ (ref.~\onlinecite{PhysRevLett.121.117002(2018)}), where the EPI is significantly enhanced by long-range Coulomb interactions, $\lambda$ of EPI is approximately 1.3.

Based on the quasiparticle spectra measured by ARPES,  we can extract the complex self-energy.
The real part of the self-energy is obtained by Re~$\Sigma(E)$ = $[\emph{k}(E) - \emph{k}_{0}(E)] \times |\emph{v}_F^{0}|$, where $\emph{k}_{0}(E)$ is the bare band momentum and $\emph{v}_F^{0}$ is the bare band velocity.
The imaginary part of the self-energy can be obtained by $|$Im~$\Sigma(E)|$ = $|\emph{v}_F^{0}|\times \mathrm{FWHM}$/2, where FWHM is the full-width at half-maximum of the MDC peaks.
By analyzing the kink of Ba$_{1-x}$K$_x$Mn$_2$As$_2$, the real part of the self-energy Re~$\Sigma$ peaks at approximately 50~meV (Fig.~2d), where the imaginary part of the self-energy Im~$\Sigma$ also shows a step (Fig.~2e), indicating the energy scale of the corresponding bosonic modes.
The Kramers-Kronig (KK) transform of Im~$\Sigma$ matches well with Re~$\Sigma$ below 100~meV (Fig.~2d), while that of Re~$\Sigma$ also matches Im~$\Sigma$ well after including a constant background and a quadratic term (Fig.~2e). %The constant background in the imaginary part of self-energy is due to the momentum broadening by impurity scattering.
%and electron-phonon scattering, which quickly saturates at a small energy scale beyond the energy resolution.
The quadratic term from the electron-electron scattering accounts for the deviation above 100~meV.
The KK conjugation of Im~$\Sigma$ and Re~$\Sigma$ demonstrates the self-consistency of the self-energy extraction.

The obtained Im~$\Sigma$ and Re~$\Sigma$ consistently indicate a bosonic energy scale \red{that extends from 0~meV to 120~meV and peaks at approximately 50~meV} (Figs.~2d-e).
%, which
% It may contribute to the self-energy at this energy scale, but in a much weaker coupling strength and not resolved in our study, while it is not related to the electron-boson interaction at approximately 50~meV.
\red{
%On the other hand, %the INS magnetic intensity of Ba$_{0.75}$K$_{0.25}$Mn$_2$As$_2$ is centred around 60~meV (ref.~\onlinecite{PhysRevB.95.224401(2017)}) [Fig. 2e].
%As the AFMM spectrum in BaMn$_2$As$_2$ generally follows the lineshape of INS magnetic intensity \cite{PhysRevB.84.094445},
%the energy scale of 0$\sim$120~meV matches that of INS magnetic intensity [Fig.~2e] (ref.~\onlinecite{PhysRevB.95.224401(2017)}), whose lineshape generally follows that of calculated AFMM spectrum (Supplementary Fig.~S6 and Ref.~\onlinecite{PhysRevB.84.094445}).
%The small energy offset between the centroid of the  kink structure ($\sim$50~meV) and the centroid of the AFMM magnon spectrum ($\sim$60~meV) is due to the slight doping difference, as will be shown in Fig.~3.
Given the quasiparticle self-energy, we can estimate the Eliashberg function $\alpha^2F(\omega)$ of the kink by $\alpha^2F(\omega)=\frac{\partial \mathrm{Im}\Sigma(E)}{\pi \partial{E}}|_{E=\omega}$ [see Methods], which characterizes the corresponding bosonic density of states weighted by their effective interactions with electrons \cite{10.1088/1367-2630/11/12/125005}.
As shown in Fig.~2f, the energy scale of $\alpha^2F(\omega)$ is distinct from the phonon energies below 25~meV according to the inelastic neutron scattering (INS) on Ba$_{1-x}$K$_x$Mn$_2$As$_2$ ($x$ = 0.25) (ref.~\onlinecite{PhysRevB.95.224401(2017)}), and the phonon spectra calculated for Ba$_{1-x}$K$_x$Mn$_2$As$_2$ ($x$ = 0 and 0.5) (Supplementary Section~3, \cite{KRESSE199615, PhysRevB.54.11169, TOGO20151, PhysRevLett.77.3865}), which excludes the phonons as the origin of the kink.
%  The phonons are situated below 25~meV.
Finite effect on the electronic self-energy from EPI could exist below 25~meV, but its observation is hindered by the much larger contribution from the electron boson interaction that peaks at approximately 50~meV.
Instead, the energy scale of $\alpha^2F(\omega)$  matches that of magnon spectra of Ba$_{0.75}$K$_{0.25}$Mn$_2$As$_2$ (Fig.~2f), which has been resolved by the INS magnetic intensity and well reproduced by Heisenberg-model-based simulations (Supplementary Section~4, \cite{PhysRevB.95.224401(2017),Toth_2015}). After considering the momentum restriction of single magnon absorption/emission processes,  the simulated $\alpha^2F(\omega)$ also exhibits the same energy scale (Supplementary Section~5, \cite{PhysRevB.95.224401(2017)}). %, and the simulated magnon contribution based on the approximation of single magnon absorption/emission processes, at the measured electronic momentum (Supplementary Section~5).
%Though the partial magnon intensity and $\alpha^2F(\omega)$ are slightly off in line-shape as it is determined by various factors beyond the current approximation,
%and they agree fairly well in energy scale.
%Moreover, high density of states (DOS) of magnons are expected in the Mn AFM order with spin S=5/2, which could be partly responsible for the large coupling constant $\lambda$.
The consistent energy scale provides compelling evidence that the kink feature in Ba$_{1-x}$K$_x$Mn$_2$As$_2$ is due to the strong interactions between itinerant electrons and a large range of AFMMs allowed by the energy and momentum restrictions of the corresponding scattering processes.
}

\begin{figure}[hb!]
	\centering
	\includegraphics[width=86mm]{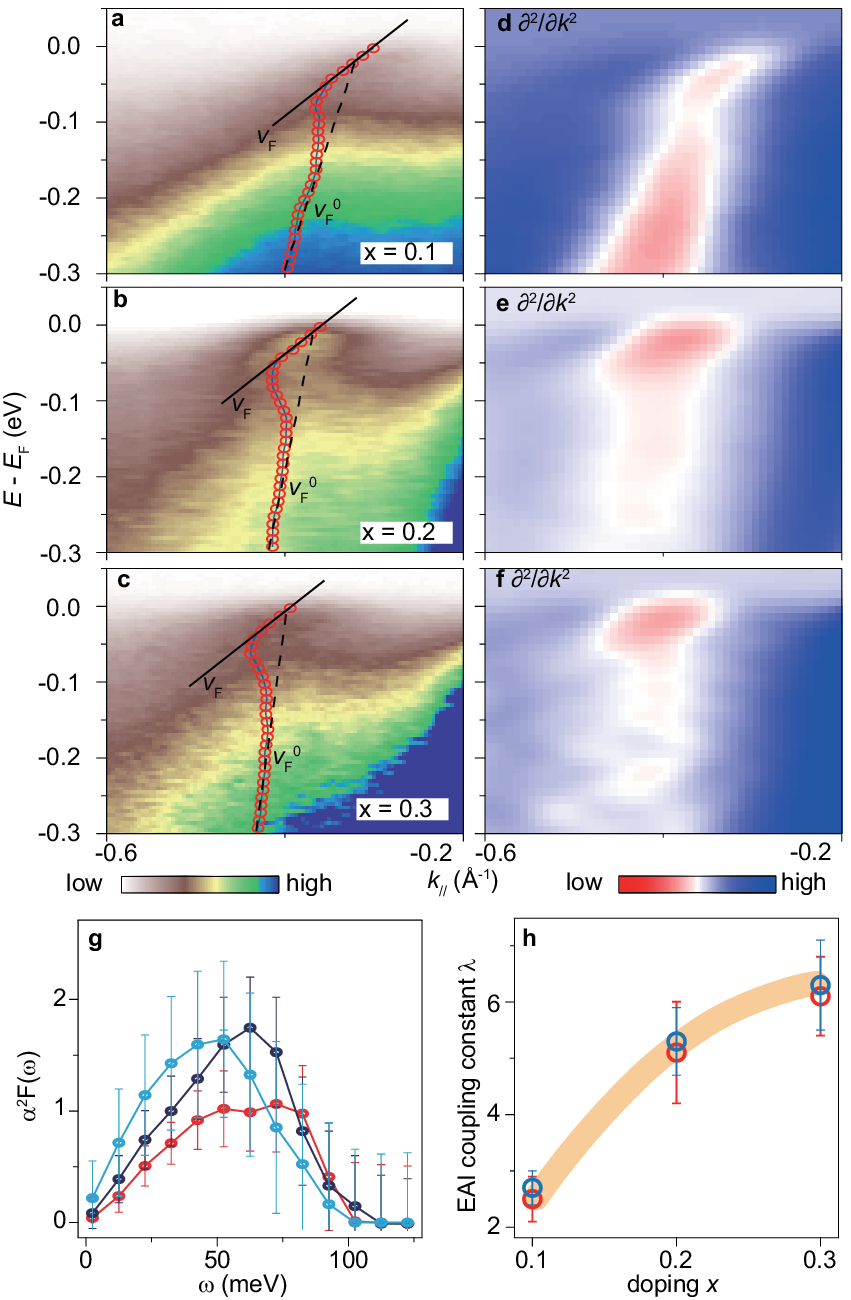}
	\caption{\textbf{Doping dependence of electron-magnon interactions.} (\textbf{a-c}) Photoemission intensity near the Fermi crossing of the $\alpha$ band along $\Gamma$X, measured with 78~eV photons at 30~K for samples of $x$ = 0.1, 0.2 and 0.3, respectively. The overlaid dispersion is obtained by fitting the MDCs. The solid lines and dashed lines illustrate the Fermi velocity and the bare band velocity, respectively. (\textbf{d-f}) Corresponding second derivative of panels (\textbf{a-c}). (\textbf{g}) \red{Eliashberg function $\alpha^2F(\omega)$ for samples of different dopings.} (\textbf{h}) Evolution of the coupling \red{constant} as a function of the doping  calculated through the band velocity renormalization (red) and Im~$\Sigma$  (blue).
	}
\end{figure}

Figures~3a-f show that kinks are present for various dopings, indicating the robust presence of electron-boson interactions in   Ba$_{1-x}$K$_x$Mn$_2$As$_2$.
Following the same procedure,  $\alpha^2F(\omega)$ is obtained for each doping.
\red{The intensity gradually shifts towards lower energy with increasing doping (Fig.~3g), which is consistent with the doping dependent energy shift of magnons as revealed by both the INS magnetic intensity and the calculated magnon DOS (Supplementary Section~4, \cite{PhysRevB.95.224401(2017),Toth_2015}). %These two doping dependencies coincide quantitatively.
%\red{As the AFMM DOS in BaMn$_2$As$_2$ generally follows the same lineshape of INS magnetic intensity \cite{PhysRevB.84.094445}, the nice match between $\alpha^2F(\omega)$ and the INS magnetic intensity justifies our assumption of no strong variation of coupling strength with the electronic energy when estimating $\alpha^2F(\omega)$.}
%Both the energy distribution and doping dependence shown in Fig.~2 and Fig.~3 provide compelling evidence that the kink feature in Ba$_{1-x}$K$_x$Mn$_2$As$_2$ is due to the strong interactions between itinerant electrons and AFMMs, \textit{i.e.} EAIs.
%The INS magnetic intensity decreases slowly with increasing doping, by approximately 10\% from $x$ = 0 to $x$ = 0.25 [Fig.~3g].
The intensity of $\alpha^2F(\omega)$ increases by nearly 50\% from $x$ = 0.1 to $x$ = 0.2, and is similar between $x$ = 0.2 and 0.3, indicating a nonlinear doping dependence of the EAIs coupling constant.
As the doping increases from $x$ = 0.1 to 0.3, the bare band velocity $v_F^0$ increases from $\sim$ 3.9 eV$\cdot \rm{\AA}$ to $\sim$ 7.7 eV$\cdot \rm{\AA}$, while the renormalized Fermi velocity $v_F$ remains at approximately $\sim$ 1.1 eV$\cdot \rm{\AA}$ (Figs.~3a-c).
Therefore, the EAIs constant $\lambda$=$\emph{v}_{F}^{0}$/$\emph{v}_{F}$-1 %,
%either estimated by   $\emph{v}_{F}^{0}$/$\emph{v}_{F}$-1 or calculated  using the Eliashberg function [see Methods],
increases with higher doping (Fig.~3h), as the system enters deep into the ferromagnetic regime.}

\section*{The mechanism of the ferromagnetic ground state}

The microscopic origin of the weak ferromagnetism in Ba$_{1-x}$K$_x$Mn$_2$As$_2$ has been under debate. \red{The picture of the canted Mn$^{2+}$ local moment \cite{PRB89.060403} was excluded by various studies, and meanwhile,  the ferromagnetism was attributed to the spin polarization of As-$4p$ holes, suggesting an itinerant picture following Stoner mechanism \cite{PhysRevLett.111.047001(2013),PhysRevLett.114.217001(2015)}.}
Currently, one strong criticism for  the Stoner mechanism is that the calculated DOS at $E_\mathrm{F}$ ($N(E_\mathrm{F})$) in Ba$_{1-x}$K$_x$Mn$_2$As$_2$ is too small to meet the Stoner criterion even at the high doping level of $x$ = 0.4 (ref.~\onlinecite{PRB89.060403}). However, the EAIs were not considered in previous  calculations, and the observed colossal EAIs would greatly enhance the DOS at $E_\mathrm{F}$. Particularly, it becomes even stronger when entering the ferromagnetic state with  increasing doping (Fig.~3h).  Therefore, the Stoner mechanism needs to be reexamined.

\begin{figure}
	\centering
	\includegraphics[width=160mm]{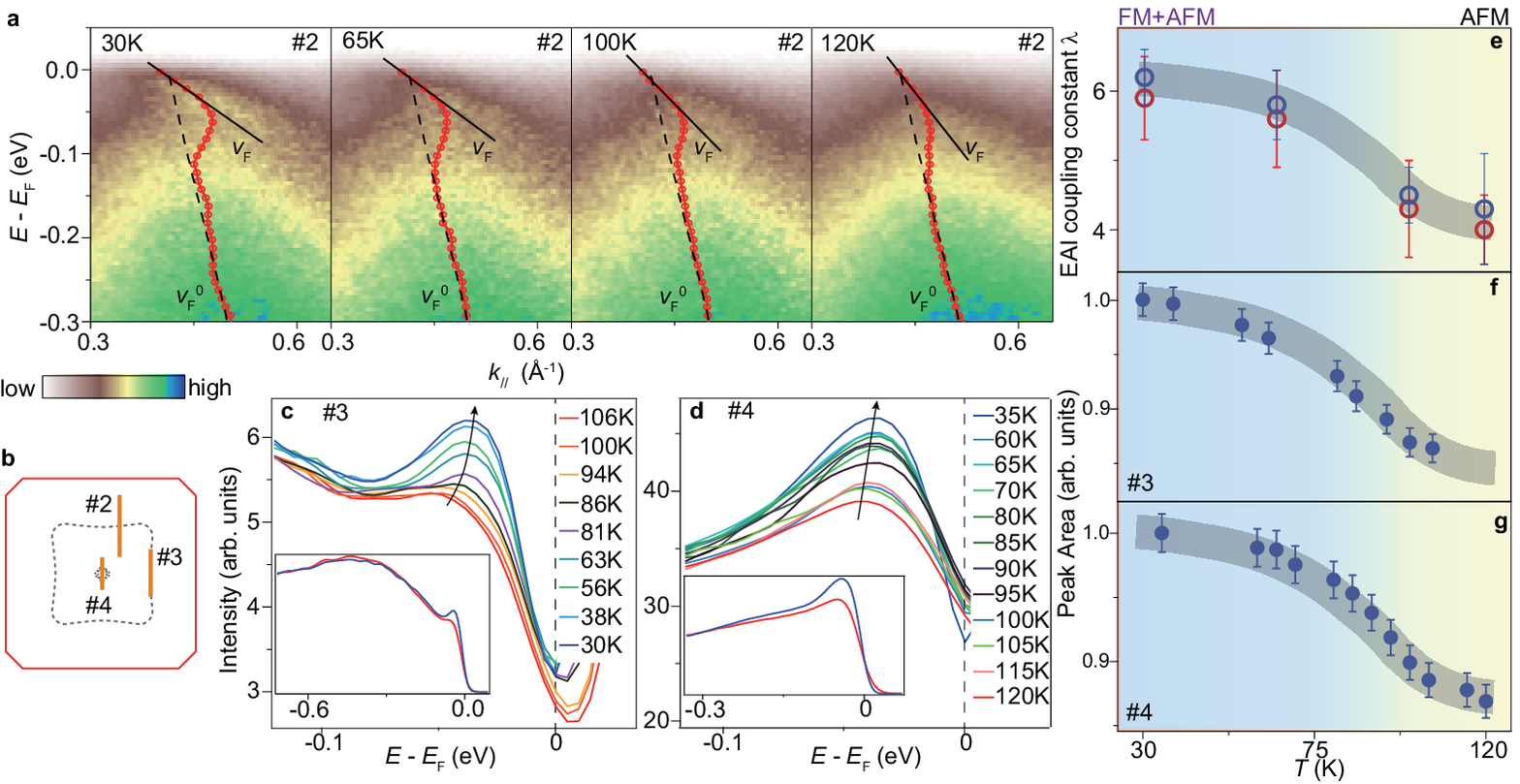}
	\caption{\textbf{Evolution of Electron-magnon interaction as a function of the temperature.} (\textbf{a}) Photoemission intensity along momentum cut~\#2 of Ba$_{1-x}$K$_x$Mn$_2$As$_2$ ($x=0.3$) at temperatures of 30~K, 65~K, 100~K and 120~K, overlaid with the band dispersion determined by the MDC peaks (red circles). The solid lines and dashed lines illustrate the Fermi velocity and the bare band velocity, respectively. (\textbf{b}) Sketch of the Fermi surface and momenta of photoemission cuts. (\textbf{c-d}) Photoemission spectra of different temperatures integrated along cut \#3 and cut \#4 in panel b, respectively. The spectra are divided by a resolution-convolved Fermi-Dirac function to remove the thermal effect near $E_\mathrm{F}$. The insets show integrated spectra at a larger energy window. (\textbf{e}) Coupling \red{constant} as a function of the temperature. The results are obtained by band velocity renormalization (red) and Im~$\Sigma$ (blue). (\textbf{f-g}) Quasiparticle peak intensity integrated over [$E_\mathrm{F}$-0.08~eV, $E_\mathrm{F}$+0.05~eV] as a function of the temperature for the spectra in panels \textbf{c} and \textbf{d}, respectively. The colour background illustrates the transition between the AFM phase and FM-AFM coexisting phase. The spectra were measured with 78~eV photons.}
\end{figure}

In addition to increasing the doping level, another route into the ferromagnetic state is through cooling. Figure~4 examines how the electronic structure of Ba$_{0.7}$K$_{0.3}$Mn$_2$As$_2$ evolves across the Curie temperature.
Figure~4a shows that the kink is always present from 30~K to 120~K, indicating that the EAIs persist in both the coexisting FM-AFM phase and the pure AFM phase, however, the Fermi velocity $\emph{v}_{F}$ increases from $\sim$ 0.55 eV$\cdot \rm{\AA}$ to $\sim$ 0.76 eV$\cdot \rm{\AA}$.
As the bare band velocity $\emph{v}_{F}^0$ is almost fixed around 3.8~eV$\cdot\rm{\AA}$, the EAIs coupling \red{constant} shows a prominent increase of approximately 50\% from 120~K to 30~K (Fig. 4e), \red{and also confirmed by the self-energy analysis (Supplementary Section~6)}. %The calculation from Im~$\Sigma$  following the Eliashberg function gives similar results [Fig. 4e and Supplementary Figure 3].
%This outcome suggests that the DOS will increase due to EAI induced renomalization when entering the FM state. Consistently,
Meanwhile, as the temperature decreases, a quasiparticle peak emerges near $E_\mathrm{F}$ for both  $\alpha$ and  $\beta$ bands (Figs. 4c-d).
It becomes sharper at lower temperature due to enhanced coherence in the ferromagnetic phase \cite{PhysRevB.63.094415(2001),PhysRevB.69.132411(2004),Zhangeaao6791(2018)}.
We characterize the emergence of the quasiparticle peak by its spectral weight, which increases continuously with decreasing temperature. The major enhancement taking place between 90~K and 60~K (Figs. 4f-g), where the broad ferromagnetic transition occurs (Supplementary Section~7). The temperature dependence of the spectral weight follows that of $\lambda$, indicating the intimate relationship between the ferromagnetic state and the EAIs.

\begin{figure}[hb]
	\centering
	\includegraphics[width=86mm]{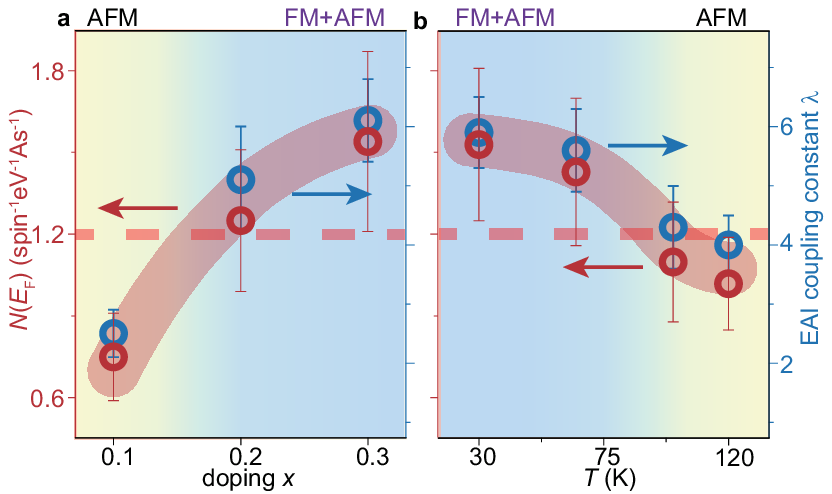}
	\caption{\textbf{Density of states and coupling \red{constant} in the temperature-doping phase diagram.} (\textbf{a}) Electronic density of states ($N(E_\mathrm{F})$, red circles) and EAIs coupling \red{constant} ($\lambda$, blue circles) as a function of the doping at 30~K. (\textbf{b}) Same as (\textbf{a}) but as a function of the temperature for $x$=0.3. The colour background illustrates the transition between the AFM phase and FM-AFM coexisting phase. The horizontal dashed line illustrates the Stoner criteria assuming $I_{\mathrm{As}}\sim$0.8~eV, to fit the phase boundary of the ferromagnetic phase.
 }
\end{figure}

The EAIs in Ba$_{1-x}$K$_x$Mn$_2$As$_2$ could influence the ferromagnetic state through the DOS at $E_\mathrm{F}$, $N(E_\mathrm{F})$.
Considering no EAIs, the bare band DOS $N^0(E_\mathrm{F})$ is approximately 0.2~spin$^{-1}$eV$^{-1}$As$^{-1}$ calculated from the measured Fermi surface and bare band velocity [see methods], which is comparable to that estimated by DFT calculations and does not meet the Stoner criteria \cite{PRB89.060403}.
After considering the EAIs, $N(E_\mathrm{F})$ is greatly enhanced from $N^0(E_\mathrm{F})$, following the doping dependence and temperature dependence of  EAIs coupling \red{constant} $\lambda$ (Fig.~5). $N(E_\mathrm{F})$ becomes most enhanced in the ferromagnetic regime, reaching 1.53$\pm$0.36~spin$^{-1}$eV$^{-1}$As$^{-1}$ at $x=0.3$, which is 3 times of that estimated by DFT calculations at an even higher doping $x$ = 0.4 (ref.~\onlinecite{PRB89.060403}).
%Both the doping and temperature dependencies of $N(E_\mathrm{F})$ are summarized in Fig.~5.
%The strong EAI renormalizes the Fermi velocity in Ba$_{1-x}$K$_x$Mn$_2$As$_2$ and the $N(E_\mathrm{F})$ is enhanced significantly, especially in the ferromagnetic regime [Figs. 5a-b].
Clearly, when $N(E_\mathrm{F})$ is roughly above $\sim$1.2~spin$^{-1}$eV$^{-1}$As$^{-1}$, Ba$_{1-x}$K$_x$Mn$_2$As$_2$ enters the ferromagnetic phase.
%As a function of the doping, the  critical doping is approximately $x$ = 0.2 at 30~K, while as a function of the temperature, the Curie temperature $T_C$ is approximately 90~K for Ba$_{0.7}$K$_{0.3}$Mn$_2$As$_2$.
If this corresponds to the Stoner criteria, $I_{\mathrm{As}}N(E_\mathrm{F})\geq1$, the Stoner parameter of As, $I_{\mathrm{As}}$, is estimated to be approximately  0.8~eV for Ba$_{0.7}$K$_{0.3}$Mn$_2$As$_2$ with a Curie temperature $T_C$ of 90~K.
We note that  the Stoner picture based on mean field theory usually fails to predict the $T_C$ of transition-metal ferromagnets with strong local exchange splitting above $T_C$ (ref.~\onlinecite{Z.Physik.B.29.34(1979)}), however, a crude estimation of $T_C$ by the Stoner criteria could still be feasible on low-$T_C$ weak ferromagnets \cite{fazekas1999lecture}, as Ba$_{1-x}$K$_x$Mn$_2$As$_2$ is supposed to be.
Consistently at 30 K, with this $I_{\mathrm{As}}\sim$0.8~eV,  the Stoner criteria are fulfilled just at doping $x$ = 0.2, whose $T_C$ is approximately 30~K, while not fulfilled for the doping $x$ = 0.1 with no ferromagnetic ground state.
These results show that the Stoner mechanism is at work and that the ferromagnetic ground state is driven by the As-$4p$ holes renormalized by EAIs.

%As a comparison, the bare band DOS $N^0(E_\mathrm{F})$ is obtained based on the measured Fermi surface and bare band velocity without considering the EAI [Figs. 5a-b].
%As the Fermi surface and bare band velocity show negligible change with temperature, the calculated $N^0(E_\mathrm{F})$ is nearly independent of temperature. The doped holes expands the Fermi surfaces while increases the bare band velocity of Ba$_{1-x}$K$_x$Mn$_2$As$_2$, resulting in nearly constant $N^0(E_\mathrm{F})$ for the measured dopings.
%For all the measured dopings and temperatures, the estimated bare $N^0(E_\mathrm{F})$ is $\sim$0.2~spin$^{-1}$eV$^{-1}$As$^{-1}$, which does not meet the Stoner criteria  \cite{PRB89.060403}.
%In contrast, after considering the EAI, the $N(E_\mathrm{F})$ is strongly dependent on temperature and doping, and greatly enhanced at low temperatures and high dopings. The estimated $N(E_\mathrm{F})$ of Ba$_{0.7}$K$_{0.3}$Mn$_2$As$_2$  at $T$=30~K is 1.53$\pm$0.36~spin$^{-1}$eV$^{-1}$As$^{-1}$, which is 3 times of that estimated by DFT calculations at an even higher doping $x$ = 0.4 (ref.~\onlinecite{PRB89.060403}). This observation again highlights the critical DOS enhancement effect of the EAI, which drives the ferromagnetic transition.

In a Stoner ferromagnet, exchange band splitting will occur.
%\red{The Stoner mechanism in Ba$_{1-x}$K$_x$Mn$_2$As$_2$ does not manifest itself along a textbook-like itinerant ferromagnet picture. As an extreme case of itinerant ferromagnetism, a half metal state was proposed for Ba$_{1-x}$K$_x$Mn$_2$As$_2$ \cite{PhysRevLett.111.047001(2013)}, but in our data the evolution of Fermi surface volume with doping deviates from the half metal behavior [Supplementary Fig.~S11].}
%, as it would have evolved twice as fast in a half metal.
%Moreover, there is no detectable splitting between a majority band and a minority band across the Curie  temperature  [Fig.~4a].
Considering the finite resolution, the upper limit of exchange splitting is estimated by fitting the MDC peaks of band $\alpha$ with two Lorentzian peaks, which gives a momentum splitting of 0.065~\rm{\AA}$^{-1}$ in Ba$_{1-x}$K$_x$Mn$_2$As$_2$ ($x$ = 0.3) at 30~K (Supplementary Section~8).
Assuming an isotropic splitting over the Fermi surface, this splitting corresponds to a spin polarization of 0.07~holes/f.u. (f.u. = formula unit), which accounts for only 32\% of the ferromagnetic moment of 0.25~$\mu_B$/f.u. at 30~K by magnetic susceptibility measurements (Supplementary Section~7).
\red{Despite that Stoner picture is still at work in driving the ferromagnetism,} at least 68\% of the ferromagnetic moments are not from itinerant carriers.
Since the magnetic moment is found to be of As character based on previous X-ray dichroism measurements  \cite{PhysRevLett.114.217001(2015)}, our results imply that the non-magnetic element As gives a considerable contribution of local moments. Magnetic moments on non-magnetic atoms could be induced by local environments as exemplified by the V moments observed in VAu$_4$ (ref.~\onlinecite{PhysRevLett.18.851(1967),doi:10.1143/JPSJ.48.62}).
In Ba$_{1-x}$K$_x$Mn$_2$As$_2$, projection of the As-$4p$ orbital was found both at the dispersive bands ($\alpha$, $\beta$, and $\beta'$), and at the flat bands at higher binding energies with localized nature \cite{PhysRevB.94.155155(2016)}. Driven by the spontaneous spin polarization of the itinerant As-$4p$ holes, the As-$4p$ states with both itinerant and localized nature in Ba$_{1-x}$K$_x$Mn$_2$As$_2$ can host an emergent ferromagnetic state with mixed itinerant and localized As moments.

\section*{Summary and Outlook}
The EAI is one of the fundamental interactions in condensed matter.
\red{Our findings  illustrate quantitatively  how the EAIs can renormalize the electronic band, invoke the Stoner mechanism, and  induce ferromagnetism in Ba$_{1-x}$K$_x$Mn$_2$As$_2$.}
It offers an unprecedented opportunity  to directly study  the self-energy and Eliasgberg function $\alpha^2F(\omega)$ of EAIs, and our quantitative analysis reveals various important facts about the EAIs  for the first time.

\begin{enumerate}
\item The coupling constant $\lambda$ of EAIs in Ba$_{1-x}$K$_x$Mn$_2$As$_2$ is as high as 6, which is much higher than any other known electron-boson coupling in metals. \red{High density of states (DOS) of magnons are expected in the Mn AFM order with spin $S$=5/2, which could be partly responsible for the large coupling constant $\lambda$}. The large coupling constant demonstrates that the effect of the EAIs can be overwhelmingly strong and could dominate the ground states.

%\item \red{Despite of the consistent energy scale between the extracted $\alpha^2F(\omega)$ and the simulated magnon contribution, the line-shape cannot be reproduced (Supplementary Section~6). Moreover, no obvious momentum dependence of the kink has been observed (Supplementary Fig.~S6). Other factors beyond the approximation of single magnon process may play significant role in the line-shape and momentum dependence of $\alpha^2F(\omega)$, e.g. the momentum and energy dependencies of the electron-magnon coupling function $g$ or the higher order multi-magnon processes (which is likely dominant considering the large λ).}

\item The EAIs vary strongly with both temperature and doping, even if the AFM order is robust. In the case of in Ba$_{1-x}$K$_x$Mn$_2$As$_2$, the EAIs is enhanced by nearly 150\% from $x$ = 0.1 to 0.3, which could be induced by the expanded Fermi surface volume and the increased scattering phase space at higher dopings. %possibly because higher doping expands the Fermi surface and increases the scattering phase space.
    This observation is in contrast to the EPI, which usually exhibits an opposite behavior.
\end{enumerate}

\red{
Our data put constrains on the theories of EAIs.
The extracted electronic self-energy and Eliashberg function $\alpha^2F(\omega)$ will motivate future inelastic neutron scattering studies to obtain the magnon spectra with absolute value, so that the electron-magnon matrix element $g$ and other microscopic characteristics of EAIs could be further revealed at a more quantitative level.
As these characteristics of the EAIs could be general in other materials with AFM excitations, such as high temperature superconductivity in cuprates and iron pnictides, our results would facilitate further understanding of the related emergent phenomena.}
%these findings  provide a good starting point for further  establishment of  theories for understanding other emergent phenomena, such as high temperature superconductivity, in correlated materials with strong EAIs.

\section*{Methods}
\subsubsection{Sample growth and characterization.}
The Ba$_{1-x}$K$_x$Mn$_2$As$_2$ single crystals used in this study were grown by the flux method as described elsewhere \cite{PhysRevB.85.144523(2012),PRB79.075120}. The chemical composition and K doping level were determined by electron probe microanalysis (EPMA).
\red{The carrier concentrations calculated from Fermi surface volumes based on Luttinger theorem agree well with the chemical doping determined by EPMA measurements, indicating that the Ba$_{1-x}$K$_x$Mn$_2$As$_2$ sample is of single phase with homogeneous doping level. This is further supported by STM and EPMA measurements over the cleaved surface (Supplementary Section~2, \cite{PhysRevB.94.155155(2016), nphys1908, PhysRevB.87.144418(2013)}). %Besides, the band structure is of single set, with no overlapping of the photoemission spectra from BaMn2As2 on doped Ba$_{1-x}$K$_x$Mn$_2$As$_2$. These results indicate that the sample is homogeneously doped and the AFM is from Ba$_{1-x}$K$_x$Mn$_2$As$_2$ itself [More characterization on the sample homogeneity is shown in Supplementary Section~1].
}
The magnetic susceptibility was measured by Quantum Design Dynalcool system (Supplementary Section~7).

\subsubsection{ARPES experiments.}
VUV-ARPES experiments were performed at beamlines BL5-2 of the Stanford Synchrotron Radiation Light Source (SSRL), I05 of the Diamond Light Source, and BL4.0.3 of the Advanced Light Source (ALS). The energy and angular resolutions were set at 15~meV and 0.3$^{\circ}$, respectively. SX-ARPES experiments were performed at ADRESS of the Swiss Light Source \cite{strocov2014soft}. The energy and angular resolutions were set at 80~meV and 0.1$^{\circ}$, respectively. All samples were cleaved \emph{in situ} under a vacuum better than 2$\times$10$^{-10}$ mbar and measured under a vacuum better than 8$\times$10$^{-11}$ mbar.

\subsubsection{Extraction of self-energy and calculation of the coupling \red{constant}.}
The real part of the self-energy can be obtained by Re~$\Sigma(E,k)$ = $E(k) - \epsilon(k) = (\emph{k}(E) - \emph{k}_{0}(E)) \times |\emph{v}_F^{0}|$, where the $\epsilon(k)$, $E(k)$, $\emph{k}_{0}(E)$, and $\emph{v}_F^{0}$ are the bare band energy, renormalized energy, bare band momentum, and bare band velocity, respectively. The bare band is estimated by cubic spline interpolation. The imaginary part of the self-energy can be derived by $|Im\Sigma(E,k)|$ = $|\emph{v}_F^{0}|\times\mathrm(FWHM)$ /2, where $\mathrm(FWHM)$ is the full-width at half-maximum of the MDC peak and $\emph{v}_F^{0}$ is the bare band velocity. The Re~$\Sigma$ and Im~$\Sigma$ should be the Kramers-Kronig conjugation of each other before the cut-off energy $E_{max}$=115~meV:

\begin{equation*}
Re \Sigma(E, k) = \frac{1}{\pi} \int_{ - \infty }^{ + \infty } {\frac{Im \Sigma(E, k)}{E' - E} dE'} ~E \leq E_{max}
\end{equation*}

\begin{equation*}
Im \Sigma(E, k) = \left\{
                                    \begin{aligned}
                                    -\frac{1}{\pi} \int_{ - \infty }^{ + \infty } {\frac{Re \Sigma(E, k)}{E' - E} dE'} ~E \leq E_{max} & \\
                                    -\frac{1}{\pi} \int_{ - \infty }^{ + \infty } {\frac{Re \Sigma(E_{max}, k)}{E' - E_{max}} dE'} ~E > E_{max} &
                                    \end{aligned}
\right.
\end{equation*}

The cut-off energy $E_{max}$ is determined by the maximum energy of AFM magnons $\omega_{max}\sim115$~meV in Ba$_{1-x}$K$_x$Mn$_2$As$_2$, above which the electron-electron correlation starts to substantially contribute to the increase in Im~$\Sigma$.

The coupling constant is calculated by either the renormalization of the Fermi velocity, or by the Eliashberg function from the self-energy. For both cases, we employ the formula at the $T\rightarrow0$ limit because the experimental temperatures are far below the $T_N$ of Ba$_{1-x}$K$_x$Mn$_2$As$_2$.
In the quasielastic approximation where the electronic energy $E$ is much larger than the boson energy $\omega$, the quasiparticle self-energy is determined by the Eliashberg function as \cite{10.1088/1367-2630/11/12/125005}:

\begin{equation*}
Im\Sigma(E,k;T)=\pi\int_{0}^{\omega_{max}}{\alpha^2F(E,k;\omega)[1-f(E-\omega)+f(E+\omega)+2n(\omega)]}d\omega
\end{equation*}

Assuming the $T\rightarrow0$ limit with $n(\omega)\rightarrow0$, we have:

\begin{equation*}
Im\Sigma(E,k)=\pi\int_{0}^{\omega_{max}}{\alpha^2F(E,k;\omega)f(E+\omega)}d\omega
\end{equation*}

As the electronic energy scale of band $\alpha$ is much larger than the magnon energy scale, the dependence of $\alpha^2F(E,k;\omega)$ on the electronic energy can be ignored. In this case, the variation in Im$\Sigma(E)$ with $E$ mainly comes from the change in the effective integration range due to the Fermi-Dirac function $f(E+\omega)$. Thus we can estimate $\alpha^2F(\omega)$ by taking the derivative of Im$\Sigma(E)$:

\begin{equation*}
\alpha^2F(\omega)=\frac{\partial Im\Sigma(E)}{\pi \partial{E}}|_{E=\omega}
\end{equation*}

To avoid nonphysical values, we set the noise induced negative data points of $\alpha^2F(\omega)$ to zero.
The coupling \red{constant} $\lambda$ can be estimated by:

\begin{equation*}
    \lambda = 2 \int_{0}^{\omega_{max}} {\frac{{\alpha ^{2}}F(\omega)} {\omega}}d\omega
\end{equation*}

\subsubsection{Estimation of density of states \textit{N}(\textit{E}$_F$) and bare-band density of states \textit{N}$^0$(\textit{E}$_F$)}

The DOS at $E_\mathrm{F}$, $N(E_\mathrm{F})$, is estimated by the following equations based on its definition:

\begin{equation*}
    g_{n}(E)=\int_{S_{n}(E)} {\frac{dS}{4\pi^3} \frac{1}{|{\nabla}E_n(k)|} \approx \frac{S_{FS}}{4\pi^3} \overline{|v_F^{-1}|}}
\end{equation*}

\begin{equation*}
    N(0)=\frac{1}{2} \times \frac{V_0}{2} g_{n} \approx \frac{V_{0}S_{FS}}{4\pi^3} \overline{|v_F^{-1}|}
\end{equation*}

where $V_{0}$ is the volume of the unit cell, $S_{FS}$ is the Fermi surface area and $v_F$ is the Fermi velocity. The bare-band density of states $N^0(E_\mathrm{F})$ are obtained in the same way by replacing the Fermi velocity $v_F$ with the bare-band velocity $v^0_F$.

We estimate the $S_{FS}$ of the drum-shaped Fermi pocket $\alpha$ by an analogue of the body-centred tetragonal Brillouin zone, and the $S_{FS}$ of band $\beta$/$\beta'$ based on elliptical cylinders. The uncertainty of the $S_{FS}$ is estimated to be 15$\%$. The $\overline{v_F^{-1}}$ calculation is based on the measured value of $v_F$ and an assumed linear change. Thus,

\begin{equation*}
    \overline{v_F^{-1}}=\frac{\int_{v_{Fmin}}^{v_{Fmax}} {v_F^{-1}}d{v_F}} {v_{Fmax} - v_{Fmin}} = \frac{ln(v_{Fmax})-ln(v_{Fmin})}{v_{Fmax}-v_{Fmin}}.
\end{equation*}

\bibliographystyle{naturemag}
\bibliography{sample}

We gratefully acknowledge the experimental support of Dr. Y. B. Huang, Dr. Z. Sun, Dr Z. T. Liu, and Dr. D. W. Shen.
We thank the Diamond Light Source for time on beam line I05, the Advanced Light Source (U.S. DOE contract no. DE-AC02-05CH11231) for access to beamline 4.0.3,
the Stanford Synchrotron Radiation Light Source for the access to beamline 5-2, and the Swiss Light Source for time on SX-ARPES endstation of beamline ADRESS. Some preliminary data were taken at  National Synchrotron Radiation Laboratory (NSRL, China) and BL03U at Shanghai Synchrotron Radiation Facility.
This work is supported in part by the  National
Natural Science Foundation of China (Grants No. 11888101, 12074074, 11704074, 11704073, and 11790312), the National
Key R\&D Program of the MOST of China (Grants No. 2016YFA0300200 and 2017YFA0303004), Project supported by Shanghai Municipal Science and Technology Major Project (Grant No. 2019SHZDZX01), and Shanghai Rising-Star Program (20QA1401400).

[Competing Interests] The authors declare that they have no
competing financial interests.

[Correspondence] Correspondence and requests for materials
should be addressed to H.C.X. (email: xuhaichao@fudan.edu.cn) and D.L.F. (email: dlfeng@fudan.edu.cn).

\end{document}